\documentclass[12pt]{article}
\usepackage{graphicx}
\usepackage[cp1251]{inputenc}
\usepackage{epsfig}
\usepackage[english]{babel}

\date{}

\begin{document}
\setcounter{page}{1}
\pagestyle{plain}
\title{\bf{Tunable Plasmon modes in doped AA-stacked bilayer graphene}}

\author{Yawar Mohammadi \thanks{Corresponding author's E-mail address:
y.mohammadi@cfu.ac.ir}} \maketitle{\centerline{Department of
physics, Farhangian University, Tehran, Iran}

\begin{abstract}
We study plasmon modes in doped AA-stacked bilayer graphene (BLG) within the nearest-neighbor tight-binding and the random phase approximation. We obtain closed analytical expressions for the polarizability function which are used to obtain the low-energy dispersion relations of and the numerical results for both acoustic and optical plasmon modes. Our result reveal the potential of AA-stacked BLG to be used as a tunable plasmonic device. In particular we find that the long-wavelength acoustic plasmon disperse as $\omega_{+}\approx\sqrt{max(|\mu|,t_{1})q}$ with a phase space which shrinks and vanishes as the chemical potential approaches the interlayer hopping energy, preventing the existence of long-lived acoustic plasmon. Furthermore, we show that AA-stacked BLG support undamped optical plasmon only when the condition $(1+\frac{g_{\sigma}g_{v}e^{2}t_{1}d}{\kappa v_{F}^{2}}\frac{|\mu|}{t_{1}})^{1/2}<\frac{|\mu|}{t_{1}}$ is satisfied, specially indicating the optical plasmon in undoped AA-staked BLG is Landau damped even at long-wavelength limit. We also find that the optical plasmon mode disperses as $\omega_{-}\approx \Delta+Cq^{2}$ with constants that can be tuned by tuning the chemical potential.

\end{abstract}


\vspace{0.5cm}

{\it \emph{Keywords}}: A. AA-stacked bilayer graphene; D.
Random phase approximation; D. Tight-binding model; D.
plasmon.
%
\section{Introduction}
\label{sec:1}

Properties of few-layer graphene strongly depend on the number of the layers and their stacking order\cite{Castro Neto1,Das Sarma1}. Few-layer graphene are found naturally in ABA(Bernal) and ABC(rhombohedral) stacking order and are made by exfoliation from graphite \cite{Novoselov1}. They can also be synthesised by chemical vapor deposition\cite{Zhang1,Yan1} in natural and specially in AAA stacking order\cite{Liu1,Lauffer1}. Most researches focus on the properties of the few-layer graphene with natural staking order which have reach physics including anomalous quantum Hall effect\cite{Novoselov1,Kou1}, tunable band gap\cite{Zhang1}, potential for excitonic condensation\cite{Barlas1} and  many other exotic properties\cite{Zou1,Lee1,Cao1}. While, AA-stacked few-layer graphenes, due to their special staking order, have especial band structure composed of electron-, hole-doped and even undoped(for few-layer graphene with odd number of layers) linear graphene-like band structures\cite{Borysiuk1}, leading to interesting properties like metal-insulator phase transition\cite{Sboychakov1}, Spin Hall effect\cite{Dyrdal1}, tunable local moment formation\cite{Mohammadi1} and anomalous quantum Hall effect\cite{Hsu1}.

One of the most important physical properties of a physical system is the dielectric function. Obtaining this quantity is essential to study many fundamental properties\cite{Bruss1}, e.g. screened Coulomb interaction, optical conductivity, and especially collective excitations (plasmon modes). Recently several groups have studied the screening and the plasmon modes in single layer graphene\cite{Wunsch1,Hwang1,Pyatkovskiy1}, AB-stacked BLG in two\cite{Wang1,Sensarma1} and four\cite{Borghi1,Gamayun1,Triola1,Ho1} band models and also in few-layer graphene\cite{van Gelderen1,Lin1,Mohammadi2}. Plasmon modes also have been studied in AA-stacked BLG. Roldan and Brey\cite{Roldan1} showed that AA-stacked BLG exhibit both acoustic and optical plasmon modes. Their result for the optical plasmon mode (which is found to be gapless and disperse as $q$) is contrast to what was reported by Hwang and Das Sarma\cite{Hwang2} for coupled bilayer structures and also is contrast to the results of Lin and coworkers that studied analytically the plasmon modes in pristine AA-stacked BLG in the absence\cite{Chuang1} and the presence of a gate voltage\cite{Chuang2}. In this paper, we extend the previous researches and investigate the effect of doping on the plasmon modes in AA-stacked BLG. We calculate polarizability function in AA-stacked BLG within the random phase approximation (RPA) and obtain the low-energy dispersion relations of and the numerical results for both acoustic and optical plasmon modes. We study the dependence of the plasmon dispersion relation on the chemical potential. In particular we show that the long-wavelength acoustic plasmon mode disperse as $\omega_{+}\approx\sqrt{max(|\mu|,t_{1})q}$ with a phase space which shrinks and vanishes as the chemical potential approaches the interlayer hopping energy, preventing the existence of long-lived acoustic plasmons. Furthermore, we show that always there is a value for the chemical potential, determined by $(1+\frac{1.015}{\kappa}\frac{|\mu|}{t_{1}})^{1/2}=\frac{|\mu|}{t_{1}}$ condition, below which and specially in the undoped case the optical plasmon mode is landau damped, while above which AA-stacked BLG support coherent optical plasmon which disperses as $\omega_{-}\approx \Delta+Cq^{2}$ with constants that can be tuned by tuning the chemical potential. These results reveal the potential of doped AA-stacked BLG to be used as a tunable plasmonic device.

\section{Model and Formalism}
\label{sec:2}

In this section, first we introduce the tight-binding Hamiltonian of AA-stacked BLG and obtain the corresponding eigenvalue and eigenfunction. Then we obtain a Lindhard-like polarizibility function within the random phase approximation.

\subsection{Tight-binding Hamiltonian and energy spectrum}
\label{subsec:21}

An AA-stacked BLG is composed of two single-layer graphenes, in which each sublattice of the upper layer is located directly above the same one in lower layer. The unit cell of an AA-stacked BLG consists of four carbon atoms, two atoms for each layer. The corresponding Briloin zone, as in graphene, is a two-dimensional hexagon. Within the nearest-neighbor tight-binding approximation, the low-energy physics in AA-stacked BLG, which as in graphene occurs near the two Dirac points $\mathbf{K}$ and $\mathbf{K^{'}}$, is governed by a $4\times4$ Hamiltonian matrix as\cite{Castro Neto1,Roldan1}
\begin{eqnarray}
\hat{H}(\mathbf{k})=\left(
\begin{array}{c c c c}
0   &  v_{F}(k_{x}\mp ik_{y})   &  t_{1}   &  0 \\
v_{F}(k_{x}\pm ik_{y})  &  0   &  0  &  t_{1} \\
t_{1}  &  0    &  0   &  v_{F}(k_{x}\mp ik_{y}) \\
0  &  t_{1}    &  v_{F}(k_{x}\pm ik_{y})   &  0 \\
\end{array}
\right),
\label{eq01}
\end{eqnarray}
where $v_{F}\simeq 9\times10^{5} ms^{-1}$ is the Fermi velocity, $t_{1}=0.2~eV$ is the inter-layer hopping energy, and the wave vector $\mathbf{k}=(k_{x},k_{y})$ is measured with respect to $\mathbf{K}$($\mathbf{K^{'}}$) if the upper(lower) sign is used.

By diagonalizing the Hamiltonian matrix, we obtain its low-energy eigenvalues,
\begin{equation}
\varepsilon_{s,\lambda}(\mathbf{k})=st_{1}+\lambda v_{F}k,
\label{eq02}
\end{equation}
where $k=(k_{x}^{2}+k_{y}^{2})^{1/2}$, $s=\pm1$ and $\lambda=\pm1$ denotes to two decoupled energy bands and their sub-bands respectively (See Fig. \ref{Fig01}). The corresponding eigenfunctions are given by\cite{Mohammadi3}
\begin{eqnarray}
\hat{\Psi}_{s,\lambda}(\mathbf{k})=\frac{1}{2}\left(
\begin{array}{c }
1 \\
\lambda e^{\pm i\phi_{\mathbf{k}}}  \\
s \\
s\lambda e^{\pm i\phi_{\mathbf{k}}}  \\
\end{array}
\right),
\label{eq03}
\end{eqnarray}
where $\phi_{\mathbf{k}}=tan^{-1}(k_{y}/k_{x})$ and the upper(lower) sign denotes to $\mathbf{K}$($\mathbf{K^{'}}$) Dirac point.

\subsection{polarizability and dielectric function in the random-phase approximation}
\label{subsec:21}

Within the random phase approximation the dielectric function is given by
\begin{equation}
\hat{\epsilon}^{RPA}(\mathbf{q},\omega)=\hat{1}-\hat{V}(q)\hat{\Pi}^{0}(\mathbf{q},\omega),
\label{eq04}
\end{equation}
where $\hat{V}(q)$ and $\hat{\Pi}^{0}(\mathbf{q},\omega)$ are the Coulomb interaction matrix and the non-interacting polarizibility matrix respectively, which for a bilayer lattice structure are given by
\begin{eqnarray}
\hat{\Pi}^{0}(\mathbf{q},\omega)=\left(
\begin{array}{c c}
\Pi^{0}_{11}(\mathbf{q},\omega)   &  \Pi^{0}_{12}(\mathbf{q},\omega)  \\
\Pi^{0}_{21}(\mathbf{q},\omega)   &  \Pi^{0}_{22}(\mathbf{q},\omega)
\end{array}\right)
\label{eq05}
\end{eqnarray}
and
\begin{eqnarray}
\hat{V}(q)=\left(
\begin{array}{c c}
V_{11}(q)      &    V_{12}(q)   \\
V_{21}(q)      &    V_{22}(q)
\end{array}\right)=\left(
\begin{array}{c c}
\frac{2\pi e^{2}}{\kappa q}           &    \frac{2\pi e^{2}}{\kappa q}e^{-qd}    \\
\frac{2\pi e^{2}}{\kappa q}e^{-qd}    &    \frac{2\pi e^{2}}{\kappa q}
\end{array}\right),
\label{eq06}
\end{eqnarray}
where the subscripts denote to the layer number, $d=3.6$ {\AA} is the inter-layer distance, $q$ is the transformed momentum, and $\kappa$ is the background dielectric constant. For a coupled bilayer lattice structure such as  AA-stacked BLG, due to the inter-layer tunneling, all components of the polarizability matrix ,in general, are nonzero. They can be calculated by analytic continuation ($i\omega_{n}\rightarrow \omega=i\eta$) of the density-density correlation function,
\begin{equation}
\Pi_{mn}^{0}(\mathbf{q},i\omega_{n})=-\frac{1}{A}\int_{0}^{\beta} d\tau
e^{i\omega_{n}\tau} \langle T_{\tau}
\rho_{m}(\mathbf{q},\tau)\rho_{n}(-\mathbf{q},0)\rangle_{0},
\label{eq07}
\end{equation}
where $i\omega_{n}$ are bosonic Matsubara frequencies and $\rho_{m}$ is the particle density at $m^{th}$ layer. We use Wick's theorem to rewrite the polarizability matrix in terms of the single-particle Green's function matrix of the AA-stacked BLG\cite{Roldan2},
\begin{eqnarray}
\hat{\Pi}^{0}(\mathbf{q},i\omega_{n})=\frac{g_{\sigma}g_{v}}{\beta A}\sum_{\mathbf{k}}\sum_{ip_{n}}
Tr[\hat{G}^{0}(\mathbf{k}+\mathbf{q},ip_{n}+i\omega_{n})\hat{G}^{0}(\mathbf{k},ip_{n})],
\label{eq08}
\end{eqnarray}
where $g_{\sigma}$ and $g_{v}$ are the spin and valley degeneracies respectively, $A$ denotes to the area of the AA-stacked BLG, $Tr\hat{A}$ is the trace of $\hat{A}$, and $ip_{n}$ is a fermionic Matsubara frequency. After a straightforward calculation, we obtain the following equations for the components of the polarizability matrix:
\begin{eqnarray}
\Pi_{11}^{0}(\mathbf{q},i\omega_{n})=\Pi_{22}^{0}(\mathbf{q},i\omega_{n})=\frac{g_{\sigma}g_{v}}{4A}\sum_{\mathbf{k}}\sum_{s s^{'}\lambda\lambda^{'}}\frac{f_{s,\lambda}(\mathbf{k})
-f_{s^{'},\lambda^{'}}(\mathbf{k}+\mathbf{q})}
{\varepsilon_{s,\lambda}(\mathbf{k})-\varepsilon_{s^{'},\lambda^{'}}(\mathbf{k}+\mathbf{q})+i\omega_{n}}
\frac{1+\lambda\lambda^{'}\cos\phi_{\mathbf{k},\mathbf{k+q}}}{2},
\label{eq09}
\end{eqnarray}
\begin{eqnarray}
\Pi_{12}^{0}(\mathbf{q},i\omega_{n})=\Pi_{21}^{0}(\mathbf{q},i\omega_{n})=\frac{g_{\sigma}g_{v}}{4A}\sum_{\mathbf{k}}\sum_{s s^{'}\lambda\lambda^{'}}ss^{'}\frac{f_{s,\lambda}(\mathbf{k})
-f_{s^{'},\lambda^{'}}(\mathbf{k}+\mathbf{q})}
{\varepsilon_{s,\lambda}(\mathbf{k})-\varepsilon_{s^{'},\lambda^{'}}(\mathbf{k}+\mathbf{q})+i\omega_{n}}
\frac{1+\lambda\lambda^{'}\cos\phi_{\mathbf{k},\mathbf{k+q}}}{2},
\label{eq10}
\end{eqnarray}
where $f_{s,\lambda}(\mathbf{k})=[exp[\beta(\varepsilon_{s,\lambda}(\mathbf{k})-\mu)]+1]^{-1}$ is  Fermi-Dirac distribution function and $\phi_{\mathbf{k},\mathbf{k+q}}$ is the angle between $\mathbf{k}$ and $\mathbf{k+q}$ vectors. One can easily check that at $t_{1}\rightarrow0$ limit, in accordance with the well known result for the decoupled bilayer lattice structures\cite{Hwang2}, $\Pi_{12}^{0}(\mathbf{q},i\omega_{n})$ becomes zero, and $\Pi_{11}^{0}(\mathbf{q},i\omega_{n})$ becomes equal to the polarizability function of graphene. Details of calculations of $\Pi_{mn}^{0}(\mathbf{q},i\omega_{n})$ has been presented in the appendix.

\section{Results and Discussion}
\label{sec:3}

In this section we study the dispersion relation of the plasmon modes in doped AA-stacked BLG and its tunability via a gate voltage. The dispersion relation of the plasmon modes are the zeros of $|\hat{\epsilon}^{RPA}(\mathbf{q},\omega)|=0$. One can easily solve the determinant $|\hat{\epsilon}^{RPA}(\mathbf{q},\omega)|=0$ (or first apply the unitary transformation proposed in Ref. [31] and then solve the determinant) to arrive at the following equations for the plasmon modes:
\begin{eqnarray}
\epsilon_{+}^{RPA}(\mathbf{q},\omega)=1-V_{+}(q)\Pi_{+}^{0}(\mathbf{q},\omega)=0,\label{eq11}
\\
\epsilon_{-}^{RPA}(\mathbf{q},\omega)=1-V_{-}(q)\Pi_{-}^{0}(\mathbf{q},\omega)=0,
\label{eq12}
\end{eqnarray}
where $V_{\pm}(q)=\frac{2\pi e^{2}}{\kappa q}(1\pm e^{-qd})$ and $\Pi_{\pm}^{0}(\mathbf{q},\omega)$ defined as $\Pi_{\pm}^{0}(\mathbf{q},\omega)=\Pi_{11}^{0}(\mathbf{q},\omega)\pm\Pi_{12}^{0}(\mathbf{q},\omega)$ which, after inserting Eqs. \ref{eq09} and \ref{eq10} into them, can be written as
\begin{eqnarray}
\Pi_{\pm}^{0}(\mathbf{q},\omega)=\frac{g_{\sigma}g_{v}}{2A}\sum_{\mathbf{k}}\sum_{s\lambda\lambda^{'}}
\frac{f_{s,\lambda}(\mathbf{k})-f_{\pm s,\lambda^{'}}(\mathbf{k}+\mathbf{q})}
{\varepsilon_{s,\lambda}(\mathbf{k})-\varepsilon_{\pm s,\lambda^{'}}(\mathbf{k}+\mathbf{q})+\omega+i\eta}
\frac{1+\lambda\lambda^{'}\cos\phi_{\mathbf{k},\mathbf{k+q}}}{2}.
\label{eq13}
\end{eqnarray}
where $\eta$ is the energy width due to deexcitation mechanisms.

As it is usual in bilayer systems and it is evident from Eqs. \ref{eq11} and \ref{eq12}, AA-stacked BLG exhibits two plasmon modes. The first one, obtained from Eq. \ref{eq11}, is an acoustic plasmon mode and corresponds to a collective excitation in which the charge densities in the two layers fluctuate in phase. The second plasmon mode (optical plasmon), obtained from Eq. \ref{eq12}, accounts for an out-of-phase charge-density oscillation in the two layers. Furthermore, it is clear from Eq. \ref{eq13} that only single particle excitations between bands with the same(different) s index (called intra-pair(inter-pair) single particle excitations) contribute to $\Pi_{+}^{0}(\mathbf{q},\omega)$($\Pi_{-}^{0}(\mathbf{q},\omega)$) polarizability. This, as we show in the following subsection, cause the acoustic and optical plasmon modes to appear in different regions of the $\omega-q$ space and show different dispersion relation. This can also lead to completely different single particle continuum for $\Pi_{+}^{0}(\mathbf{q},\omega)$ and $\Pi_{-}^{0}(\mathbf{q},\omega)$, determining the region in which the optical and the acoustic plasmon modes are damped or undamped. The
boundaries of the corresponding dissipative single particle continuum can be determined from the shape of the Fermi surface and the corresponding dispersion relation given in Eq. \ref{eq02}. Figure \ref{Fig01} (Figures \ref{Fig02} and \ref{Fig03}) show all possible single particle excitations and the corresponding single particle continuum for $\Pi_{+}^{0}(\mathbf{q},\omega)$ ($\Pi_{-}^{0}(\mathbf{q},\omega)$ for $|\mu|<t_{1}$ and $|\mu|>t_{1}$) respectively. Notice that the inter-(intra-)band regions denotes to the possible intra-pair and inter-pair single particle excitations that occur between the energy bands with same(different) $\lambda$ indices, $(s,\lambda)\rightarrow(s^{'},\lambda)~~~ ((s,\lambda)\rightarrow(s^{'},-\lambda))$.

Utilizing the method of Ref. [37], we obtain analytical expressions for the real and imaginary parts of $\Pi_{+}^{0}(\mathbf{q},\omega)$ and $\Pi_{-}^{0}(\mathbf{q},\omega)$ in the regions of the $q-\omega$ space where the acoustic(optical) plasmon mode appears. These analytical expressions are used to calculate the plasmon modes numerically and also to obtain analytical expressions for their low-energy dispersion relations.

First we present our analytical results for the real and imaginary parts of $\Pi_{+}^{0}(\mathbf{q},\omega)$ and obtain the dispersion relation of the acoustic plasmon mode. It is clear from Eq. \ref{eq13} that $\Pi_{+}^{0}(\mathbf{q},\omega)$ is sum of the polarization functions of two doped SLGs with different chemical potentials, $\mu+t_{1}$ and $\mu-t_{1}$. Keeping in mind the properties of the polarization function of SLG, one can conclude that the regions in which Eq. \ref{eq11} is satisfied and consequently the acoustic plasmon mode emerges includes three distinct regions denoted by I, II and III in Fig. \ref{Fig04} (For more information see Fig. \ref{Fig01}). Inside region I where $v_{F}q<\omega<-v_{F}q+2|\mu-t_{1}|$, our results for the real and imaginary parts of $\Pi_{+}^{0}(\mathbf{q},\omega)]$ are
\begin{eqnarray}
Re[\Pi_{+}^{0}(\mathbf{q},\omega)]=&-&\frac{g_{\sigma}g_{v}(|\mu+t_{1}|+|\mu-t_{1}|)}{4\pi v_{F}^{^{2}}}  \nonumber  \\ &+&\frac{g_{\sigma}g_{v}q^{2}}{32\pi \sqrt{\omega-v_{F}^{2}q^{2}}} [G_{>}(\frac{2|\mu-t_{1}|+\omega}{v_{F}q})-G_{>}(\frac{2|\mu-t_{1}|-\omega}{v_{F}q}) \nonumber  \\
&&~~~~~~~~~~~~~~~~~~~+G_{>}(\frac{2|\mu+t_{1}|+\omega}{v_{F}q})-G_{>}(\frac{2|\mu+t_{1}|-\omega}{v_{F}q})] ,
\label{eq14}
\end{eqnarray}
 where $G_{>}(x)=x\sqrt{x^{2}-1}-\cosh^{-1}(x)~~|x|>1$ and $Im[\Pi_{+}^{0}(\mathbf{q},\omega)]=0$. For $v_{F}q<\omega<v_{F}q+2|\mu-t_{1}|$ and $-v_{F}q+2|\mu-t_{1}|\omega<-v_{F}q+2|\mu+t_{1}|$ (region II) we find that the real and imaginary parts of $\Pi_{+}^{0}(\mathbf{q},\omega)$ can be expressed as
\begin{eqnarray}
Re[\Pi_{+}^{0}(\mathbf{q},\omega)]=&-&\frac{g_{\sigma}g_{v}(|\mu+t_{1}|+|\mu-t_{1}|)}{4\pi v_{F}^{^{2}}}  \nonumber  \\ &+&\frac{g_{\sigma}g_{v}q^{2}}{32\pi \sqrt{\omega-v_{F}^{2}q^{2}}} [G_{>}(\frac{2|\mu-t_{1}|+\omega}{v_{F}q}) \nonumber  \\
&&~~~~~~~~~~~~~~~~~~~+G_{>}(\frac{2|\mu+t_{1}|+\omega}{v_{F}q})-G_{>}(\frac{2|\mu+t_{1}|-\omega}{v_{F}q})] ,
\label{eq15}
\end{eqnarray}
and $Im[\Pi_{+}^{0}(\mathbf{q},\omega)]=-\frac{g_{\sigma}g_{v}q^{2}}{32\pi \sqrt{\omega^{2}-v_{F}^{2}q^{2}}}[\pi+G_{<}(\frac{\omega-2|\mu-t_{1}|}{v_{F}q})]$ where $G_{<}(x)=x\sqrt{1-x^{2}}-\cos^{-1}(x),~~|x|<1$. Finally, for $v_{F}q<\omega<v_{F}q+2|\mu-t_{1}|$ and $\omega>-v_{F}q+2|\mu+t_{1}|$ (namely inside region III of Fig. \ref{Fig04}) we obtain
\begin{eqnarray}
Re[\Pi_{+}^{0}(\mathbf{q},\omega)]=&-&\frac{g_{\sigma}g_{v}(|\mu+t_{1}|+|\mu-t_{1}|)}{4\pi v_{F}^{^{2}}}  \nonumber  \\ &+&\frac{g_{\sigma}g_{v}q^{2}}{32\pi \sqrt{\omega-v_{F}^{2}q^{2}}} [G_{>}(\frac{2|\mu-t_{1}|+\omega}{v_{F}q})+G_{>}(\frac{2|\mu+t_{1}|+\omega}{v_{F}q})],
\label{eq16}
\end{eqnarray}
and $Im[\Pi_{+}^{0}(\mathbf{q},\omega)]=-\frac{g_{\sigma}g_{v}q^{2}}{32\pi \sqrt{\omega^{2}-v_{F}^{2}q^{2}}}[2\pi+G_{<}(\frac{\omega-2|\mu-t_{1}|}{v_{F}q})+G_{<}(\frac{\omega-2|\mu+t_{1}|}{v_{F}q})]$ for the real and imaginary parts of the $\Pi_{+}^{0}(\mathbf{q},\omega)]$. One can check that our results for  $\Pi_{+}^{0}(\mathbf{q},\omega)$ at $t_{1}=0$ limit reduces to the results of Ref. \cite{Hwang3}, which is sum of the polarization function of two SLGs with chemical potentials $\mu$. The numerical solution of the acoustic plasmon mode obtained From Eqs. \ref{eq11}, \ref{eq14}, \ref{eq15} and \ref{eq16} is shown by solid red line in Fig. \ref{Fig04}. In this figure, we also show the single-particle excitation
regions in which the plasmon modes is expected to be Landau damped. Due to the Pauli exclusion principle(which Leads to $Im[\Pi_{+}^{0}(\mathbf{q},\omega)]=0$), the inter-band single-particle excitations open a gap in the long wavelength (region I) satisfying $v_{F}q<\omega<2|\mu-t_{1}|-v_{F}q$, in which the
coherent plasmon modes exist. As $q$ increases the mode inter region II and decay by producing interband electron-hole pair in $s=+$ band(See Fig. \ref{Fig01}). As $q$ increases further the mode inter into region III and decay due to particle-hole excitations in both $s=+$ and $s=-$ bands. Another note of worth mentioning is that, according to Fig. \ref{Fig01} which shows the boundaries of the corresponding dissipative electron-hole continuum for $\Pi_{+}^{0}(\mathbf{q},\omega)$, when $\mu=t_{1}$ the electron-hole continuum covers the entire $(\omega,q)$-space at low frequencies. Consequently the acoustic plasmon mode becomes landau-damped even at very long-wavelength limit. Furthermore, it is evident from Fig \ref{Fig04} that the long-wavelength limit of the acoustic mode can be well approximated by a low energy dispersion relation (shown by dashed red line), obtained by inserting the long-wavelength limit of $\Pi_{+}^{0}(\mathbf{q},\omega)$ into Eq. \ref{eq11}. Our analytical result for the low energy dispersion relation of the acoustic plasmon mode,
\begin{equation}
\omega_{-}(q) \approx \sqrt{\frac{g_{\sigma}g_{v}e^{2}max(t_{1},|\mu|)}{\kappa}q},
\label{eq17}
\end{equation}
 is same as that of Ref. \cite{Roldan1} \textit{obtained by calculating Drude weight of Doped AA-stacked BLG}, indicating the $\sqrt{q}$ behavior of the acoustic plasmon dispersion which is same as that in graphene and conventional two dimensional materials. Equation \ref{eq17} also show that the characteristic low-energy dispersion relation is unaffected by tuning the chemical potential unless the chemical potential exceeds the interlayer hopping energy.

Properties of the optical plasmon mode of AA-stacked BLG depend on whether the value of the chemical potential is less or larger than the inter-layer hopping energy. First we investigate $|\mu|\leq t_{1}$ case. By calculating $q=0$ limit of the optical plasmon mode for $\mu|<t_{1}$ which results in $\omega_{-}(q=0)=2t_{1}(1+\frac{g_{\sigma}g_{v}e^{2}t_{1}d}{\kappa v_{F}^{2}}\frac{\mu^{2}+t_{1}^{2}}{2t_{1}^{2}})^{1/2}$(see appendix), one can conclude that the optical plasmon mode emerge in the region I  of Fig. \ref{Fig05}, where $\omega>v_{F}q+2t_{1}$. Consequently we only need to calculate the polarization function in this region. Our results for the real and imaginary parts of $\Pi_{-}^{0}(\mathbf{q},\omega)$ in this region are
\begin{eqnarray}
Re[\Pi_{-}^{0}(\mathbf{q},\omega)]=&-&\frac{g_{\sigma}g_{v}t_{1}}{2\pi v_{F}^{^{2}}}  \nonumber  \\ &-&\frac{g_{\sigma}g_{v}q^{2}}{32\pi \sqrt{(\omega+2t_{1})^{2}-v_{F}^{2}q^{2}}} [G_{>}(\frac{\omega+2|\mu|}{v_{F}q})+G_{>}(\frac{\omega-2|\mu|}{v_{F}q})-2G_{>}(\frac{\omega+2t_{_{1}}}{v_{F}q})] \nonumber  \\
&+& \frac{g_{\sigma}g_{v}q^{2}}{32\pi \sqrt{(\omega-2t_{1})^{2}-v_{F}^{2}q^{2}}} [G_{>}(\frac{\omega+2|\mu|}{v_{F}q})+G_{>}(\frac{\omega-2|\mu|}{v_{F}q})-2G_{>}(\frac{\omega-2t_{_{1}}}{v_{F}q})] ,\nonumber  \\
\label{eq18}
\end{eqnarray}
and $Im[\Pi_{-}^{0}(\mathbf{q},\omega)]=-\frac{g_{\sigma}g_{v}}{32 v_{F}^{2}}(\frac{v_{F}^{2}q^{2}}{\sqrt{(\omega+2t_{1})^{2}-v_{F}^{2}q^{2}}}
+\frac{v_{F}^{2}q^{2}}{\sqrt{(\omega-2t_{1})^{2}-v_{F}^{2}q^{2}}})$. \textit{Notice that when $|\mu|\leq t_{1}$ the acoustic plasmon mode is damped even at the long wave-length.} This is due to the single-particle excitation between $(s=+,\lambda=-)\rightarrow(s=-,\lambda=+)$ and $(s=-,\lambda=-)\rightarrow(s=+,\lambda=+)$ (See Fig. \ref{Fig01}).

For $\mu>t_{1}$ the optical plasmon mode emerge in $v_{F}q+2t_{1}<\omega<v_{F}q+2|\mu|$ region(inside region I and II of Fig. \ref{Fig06})). We find that inside region I ($v_{F}q+2t_{1}<\omega<-v_{F}q+2|\mu|$) the real and imaginary parts of the polarizability function is given by
\begin{eqnarray}
Re[\Pi_{-}^{0}(\mathbf{q},\omega)]=&-&\frac{g_{\sigma}g_{v}|\mu|}{2\pi v_{F}^{^{2}}}  \nonumber  \\ &+&\frac{g_{\sigma}g_{v}q^{2}}{32\pi \sqrt{(\omega+2t_{1})^{2}-v_{F}^{2}q^{2}}} [G(\frac{2|\mu|+\omega}{v_{F}q})-G(\frac{2|\mu|-\omega}{v_{F}q})] \nonumber  \\
&+& \frac{g_{\sigma}g_{v}q^{2}}{32\pi \sqrt{(\omega-2t_{1})^{2}-v_{F}^{2}q^{2}}} [G(\frac{2|\mu|+\omega}{v_{F}q})-G(\frac{2|\mu|-\omega}{v_{F}q})] ,
\label{eq19}
\end{eqnarray}
and $Im[\Pi_{-}^{0}(\mathbf{q},\omega)]=0$, resulting in an undamped optical plasmon mode. But inside region II of Fig. \ref{Fig06} where $-v_{F}q+2|\mu|<\omega<v_{F}q+2|\mu|$ and $v_{F}q+2t_{1}<\omega$ we have
\begin{eqnarray}
Re[\Pi_{-}^{0}(\mathbf{q},\omega)]=&-&\frac{g_{\sigma}g_{v}|\mu|}{2\pi v_{F}^{^{2}}}  \nonumber  \\ &+&\frac{g_{\sigma}g_{v}q^{2}}{32\pi \sqrt{(\omega+2t_{1})^{2}-v_{F}^{2}q^{2}}}G(\frac{2|\mu|+\omega}{v_{F}q}) \nonumber  \\
&+& \frac{g_{\sigma}g_{v}q^{2}}{32\pi \sqrt{(\omega-2t_{1})^{2}-v_{F}^{2}q^{2}}}G(\frac{2|\mu|+\omega}{v_{F}q}),
\label{eq20}
\end{eqnarray}
with $Im[\Pi_{-}^{0}(\mathbf{q},\omega)]=-[\frac{g_{\sigma}g_{v}q^{2}}{32\pi \sqrt{(\omega+2t_{1})^{2}-v_{F}^{2}q^{2}}}+\frac{g_{\sigma}g_{v}q^{2}}{32\pi \sqrt{(\omega-2t_{1})^{2}-v_{F}^{2}q^{2}}}][\pi+G_{<}(\frac{\omega-2|\mu|}{v_{F}q})]$ , indicating the optical plasmon mode inside region II is damped.

The numerical solution of the optical plasmon mode for $|\mu|\leq t_{1}$($|\mu|\geq t_{1}$) obtained From Eqs. \ref{eq12} and \ref{eq18} (\ref{eq19} and \ref{eq20}) are shown by solid red line in Fig. \ref{Fig05}(\ref{Fig06}). Its long-wavelength limit can be well approximated by a low energy dispersion relation (shown by dashed red line), obtained by inserting the long-wavelength limit of $\Pi_{-}^{0}(\mathbf{q},\omega)$ into Eq. \ref{eq12}. Our analytical result for the low energy dispersion relation of the optical plasmon mode for both $|\mu|<t_{1}$ and $|\mu|>t_{1}$ cases can be written as $\omega^{2}_{-}(q)\approx \Delta^{2}+Cq^{2}$(indicating its gapped nature) where $\Delta=\omega_{-}(q=0)$ is the zero-momentum optical plasmon mode given in the appendix and depending on whether the chemical potential is less or larger than the inerlayer hopping energy. At $v_{F}q\ll \Delta$ limit we have $\omega_{-}(q)\approx \Delta+\frac{C}{2\Delta}q^{2}$, where
\begin{eqnarray}
C=\frac{g_{\sigma}g_{v}e^{2}t_{1}d}{\kappa}[\frac{(t_{1}^{2}+|\mu|^{2})(3\Delta^{2}+4t_{1}^{2})}
{(\Delta^{2}-4t_{1}^{2})^{2}}+\frac{(\Delta^{2}+4t_{1}^{2})}
{2(\Delta^{2}-4t_{1}^{2})}-\frac{2|\mu|\Delta}{\Delta^{2}-4t_{1}^{2}}]q^{2},~~|\mu|\leq t_{1}
\label{eq21}
\end{eqnarray}
and
\begin{eqnarray}
C=\frac{g_{\sigma}g_{v}e^{2}t_{1}d}{\kappa}[\frac{2|\mu|t_{1}(3\Delta^{2}+4t_{1}^{2})}
{(\Delta^{2}-4t_{1}^{2})^{2}}+\frac{|\mu|(\Delta^{2}+4t_{1}^{2})}
{2t_{1}(\Delta^{2}-4t_{1}^{2})}-\frac{2t_{1}\Delta}{\Delta^{2}-4t_{1}^{2}}]q^{2} ,~~|\mu|\geq t_{1}.
\label{eq22}
\end{eqnarray}
One can easily recover the out-of-phase plasmon dispersion in the decoupled double-layer graphene (Eq. 4b of Ref. [36] with $k_{F1}=k_{F2}=t_{1}/v_{F}$)  from Eqs. \ref{eq22} and A.12 combined with $\omega^{2}_{-}(q)\approx \Delta^{2}+Cq^{2}$ by setting $t_{1}=0$.

It is desired to investigate the dependence of the optical plasmon mode on the chemical potential and the background dielectric constant in details. When $|\mu|\leq t_{1}$ the optical mode always appears inside the single-particle continuum(See Figs. \ref{Fig02} and \ref{Fig05}), so the mode is damped even at long-wavelength. But the situation is different for $|\mu|<t_{1}$. When the chemical potential is larger than the interlayer hopping energy, due to the phase-space restriction, the interband single-particle continuum opens
a gap in the long wavelength region satisfying $2t_{1}+v_{F}q<\omega<2|\mu|-v_{F}q$, in which the undamped plasmon mode can exist provided the mode appears inside the gap namely $\omega_{-}(q=0)<2|\mu|$. By setting the zero-momentum optical plasmon in this equation we arrive at the following condition $(1+\frac{g_{\sigma}g_{v}e^{2}t_{1}d}{\kappa v_{F}^{2}}\frac{|\mu|}{t_{1}})^{1/2}<\frac{|\mu|}{t_{1}}$ for the coherent optical plasmon in doped AA-stacked BLG. There are two parameters in this equation, the chemical potential and the background dielectric constant, which can be changed to engineer the existence and the dispersion(See also Eq. \ref{eq22}) of the undamed optical plasmons. By keeping them as variables and substitute the value of the other parameters we arrive at $(1+\frac{1.015}{\kappa}\frac{|\mu|}{t_{1}})^{1/2}<\frac{|\mu|}{t_{1}}$. Accordingly, for $SiO_{2}/Si$ substrate($\kappa=2.5$) AA-stacked BLG support undamped optical plasmon mode provided $1.22t_{1}<|\mu|$. The condition for suspended case($\kappa=1$) and AA-stacked BLG on hBN substrate($\kappa=3.5$) is $1.63t_{1}<|\mu|$ and $1.15t_{1}<|\mu|$ respectively. These results reveal that the existence and the dispersion of the undamped plasmon modes can be tuned by tuning the chemical potential or changing the dielectric background.

\section{Summary and conclusions}
\label{sec:4}

In summary, employing the nearest-neighbor tight-binding approximation we investigated plasmon modes in doped AA-stacked within the random phase approximation. We obtained analytical expression for the polarizability function which was used to calculate both acoustic and optical plasmon modes numerically and analytically. Our results reveal the potential of doped AA-stacked BLG to be used as a tunable plasmonic device. In particular, we showed that the low-energy acoustic plasmon mode disperse as $\omega_{+}\propto\sqrt{max(|\mu|,t_{1})q}$ with a phase space which shrinks and vanishes as the chemical potential approaches the interlayer hopping energy, preventing the existence of long-lived acoustic plasmons. Furthermore, we found when the chemical potential satisfies the condition $(1+\frac{g_{\sigma}g_{v}e^{2}t_{1}d}{\kappa v_{F}^{2}}\frac{|\mu|}{t_{1}})^{1/2}<\frac{|\mu|}{t_{1}}$ AA-stacked BLG support undamped optical plasmon which disperses as $\omega_{-}\approx \Delta+Cq^{2}$ with constants that can be tuned by tuning the chemical potential. While for other values of the chemical potential(and specially in the pristine AA-stacked BLG) the optical plasmon becomes landau damped.

 \nonumber \section{acknowledgment} This work has been
supported by Farhangian University.

\appendix
\section{Polarizability Function and its zero-momentum-limit}

Starting from Eq. \ref{eq07} and using Wick's theorem\cite{Roldan2}, we write the components of the polarizability matrix in terms of the product of the components of the single-particle Green's function matrix of the AA-stacked BLG as
\begin{eqnarray}
\Pi_{mn}^{0}(\mathbf{q},i\omega_{n})=\frac{g_{\sigma}g_{v}}{\beta A}\sum_{\mathbf{k}}\sum_{ip_{n}}\sum_{X,Y=A,B}
G_{X_{m}Y_{n}}^{0}(\mathbf{k}+\mathbf{q},ip_{n}+i\omega_{n})G_{Y_{n}X_{m}}^{0}(\mathbf{k},ip_{n})
\nonumber,~~~~~~~~~~~~~~~~~~~(A.1)\label{eq:AOne}
\end{eqnarray}
where $G_{X_{m}Y_{n}}(\mathbf{k},ip_{n})$ are the components of the single-particle Green's function matrix and $X_{m}$ denotes to the $X$ sublattice in the $m^{th}$ layer of AA-stacked BLG. The components of the single-particle Green's function matrix are given by

\begin{eqnarray}
G_{A_{1}A_{1}}^{0}(\mathbf{k},ip_{n})=G_{A_{2}A_{2}}^{0}(\mathbf{k},ip_{n})=G_{B_{1}B_{1}}^{0}(\mathbf{k},ip_{n})= G_{B_{2}B_{2}}^{0}(\mathbf{k},ip_{n})=\frac{1}{4}\sum_{s,\lambda}\frac{1}{ip_{n}-\varepsilon_{s,\lambda}(\mathbf{k})}
\nonumber,~~~~~(A.2)\label{eq:ATwo}
\end{eqnarray}
\begin{eqnarray}
G_{A_{1}B_{1}}^{0}(\mathbf{k},ip_{n})=G_{A_{2}B_{2}}^{0}(\mathbf{k},ip_{n})=G_{B_{1}A_{1}}^{0\ast}(\mathbf{k},ip_{n})= G_{B_{2}A_{2}}^{0\ast}(\mathbf{k},ip_{n})=\frac{e^{-i\phi_{\mathbf{k}}}}{4}\sum_{s,\lambda}\frac{\lambda}
{ip_{n}-\varepsilon_{s,\lambda}(\mathbf{k})}
\nonumber,~(A.3)\label{eq:AThree}
\end{eqnarray}
\begin{eqnarray}
G_{A_{1}A_{2}}^{0}(\mathbf{k},ip_{n})=G_{B_{1}B_{2}}^{0}(\mathbf{k},ip_{n})=G_{A_{2}A_{1}}^{0}(\mathbf{k},ip_{n})= G_{B_{2}B_{1}}^{0}(\mathbf{k},ip_{n})=\frac{1}{4}\sum_{s,\lambda}\frac{s}{ip_{n}-\varepsilon_{s,\lambda}(\mathbf{k})}
\nonumber,~~~~(A.4)\label{eq:AFour}
\end{eqnarray}
\begin{eqnarray}
G_{A_{1}B_{2}}^{0}(\mathbf{k},ip_{n})=G_{A_{2}B_{1}}^{0}(\mathbf{k},ip_{n})=G_{B_{1}A_{2}}^{0\ast}(\mathbf{k},ip_{n})= G_{B_{2}A_{1}}^{0\ast}(\mathbf{k},ip_{n})=\frac{e^{-i\phi_{\mathbf{k}}}}{4}\sum_{s,\lambda}\frac{s\lambda}
{ip_{n}-\varepsilon_{s,\lambda}(\mathbf{k})}
\nonumber,~(A.5)\label{eq:AFive}
\end{eqnarray}
where $ip_{n}$ are fermionic frequencies, and superscript $*$ indicates that $x^{*}$ is the complex conjugate of $x$.

Due to the sublattice-symmetry of the system we have $\Pi_{11}^{0}(\mathbf{q},\omega)=\Pi_{22}^{0}(\mathbf{q},\omega)$ and $\Pi_{12}^{0}(\mathbf{q},\omega)=\Pi_{21}^{0}(\mathbf{q},\omega)$. By inserting Eqs. A.2-A.5 into Eq. A.1 and evaluating the sum over fermionic Matsubara frequencies we arrive at
\begin{eqnarray}
\Pi_{11}^{0}(\mathbf{q},i\omega_{n})=\frac{g_{\sigma}g_{v}}{4A}\sum_{\mathbf{k}}\sum_{s s^{'}\lambda\lambda^{'}}\frac{f_{s,\lambda}(\mathbf{k})
-f_{s^{'},\lambda^{'}}(\mathbf{k}+\mathbf{q})}
{\varepsilon_{s,\lambda}(\mathbf{k})-\varepsilon_{s^{'},\lambda^{'}}(\mathbf{k}+\mathbf{q})+i\omega_{n}}F_{\lambda\lambda^{'}}(\mathbf{k},\mathbf{k}+\mathbf{q}),
\nonumber,~~~~~~~~~~~~~~~~~~~~(A.6)\label{eq:ASix}
\end{eqnarray}
and
\begin{eqnarray}
\Pi_{12}^{0}(\mathbf{q},i\omega_{n})=\frac{g_{\sigma}g_{v}}{4A}\sum_{\mathbf{k}}\sum_{s s^{'}\lambda\lambda^{'}}ss^{'}\frac{f_{s,\lambda}(\mathbf{k})
-f_{s^{'},\lambda^{'}}(\mathbf{k}+\mathbf{q})}
{\varepsilon_{s,\lambda}(\mathbf{k})-\varepsilon_{s^{'},\lambda^{'}}(\mathbf{k}+\mathbf{q})+i\omega_{n}}F_{\lambda\lambda^{'}}(\mathbf{k},\mathbf{k}+\mathbf{q}),
\nonumber,~~~~~~~~~~~~~~~~~(A.7)\label{eq:ASeven}
\end{eqnarray}
where
\begin{eqnarray}
F_{\lambda\lambda^{'}}(\mathbf{k},\mathbf{k}+\mathbf{q})=\frac{1}{4}[2+\lambda\lambda^{'}
(e^{i\phi_{\mathbf{k}}}e^{-i\phi_{\mathbf{k+q}}}+e^{-i\phi_{\mathbf{k}}}e^{i\phi_{\mathbf{k+q}}})]=
\frac{1+\lambda\lambda^{'}\cos\phi_{\mathbf{k},\mathbf{k+q}}}{2},
\nonumber,~~~~~~~~~~~~~~~(A.8)\label{eq:AEight}
\end{eqnarray}
is the chirality factor. Notice to $ss^{'}$ coefficient in Eq. A.7. After applying the analytic continuation to $\Pi_{mn}(\mathbf{q},i\omega_{n})$ and setting them in the $\Pi_{\pm}^{0}(\mathbf{q},\omega)=\Pi_{11}^{0}(\mathbf{q},\omega)\pm\Pi_{12}^{0}(\mathbf{q},\omega)$ we arrive at Eq. \ref{eq13}.

By calculating $\Pi^{0}_{\pm}(\mathbf{q}=0,\omega)$ and inserting them into Eqs. \ref{eq11} and \ref{eq12} we obtain the plasmon modes at zero transform momentum, the later indicates that the optical plasmon mode in AA-stacked BLG is gapped. These results can also be used to examine (the validity of) our results for the low energy dispersion relation of the plasmon modes and to find the region in which the plasmon modes emerge. To calculate $\Pi^{0}_{\pm}(\mathbf{q}=0,\omega)$ we set $q=0$ (Notice that $F_{\lambda\lambda^{'}}(\mathbf{k},\mathbf{q}=0)=\delta_{\lambda,\lambda^{'}}$, where $\delta_{\lambda,\lambda^{'}}$ is the Kronecker delta) and then evaluate the k-integral of Eq. \ref{eq13}. So, we arrive at
\begin{eqnarray}
Re[\Pi_{-}^{0}(\mathbf{q}=0,\omega)]=\frac{g_{\sigma}g_{v}t_{1}(\mu^{2}+t_{1}^{2})}{\pi v_{F}^{2}(\omega^{2}-4t_{1}^{2})}~~~~~~~~~~~~~~~|\mu|\leq t_{1},\nonumber~~~~~~~~~~~~~~~(A.9)\label{eq:ANine}\\
Re[\Pi_{-}^{0}(\mathbf{q}=0,\omega)]=\frac{2g_{\sigma}g_{v}|\mu|t_{1}^{2}}{\pi v_{F}^{2}(\omega^{2}-4t_{1}^{2})}~~~~~~~~~~~~~~~|\mu|\geq t_{1},
\nonumber~~~~~~~~~~~~~~~(A.10)\label{eq:ATen}
\end{eqnarray}
and $Re[\Pi_{+}^{0}(\mathbf{q}=0,\omega)]=0$ for the real part of the zero-temperature polarizability functions, which can be inserted into Eqs. \ref{eq11} and \ref{eq12} to obtain the following results:
\begin{eqnarray}
\omega_{-}(q=0)=2t_{1}(1+\frac{g_{\sigma}g_{v}e^{2}t_{1}d}{\kappa v_{F}^{2}}\frac{\mu^{2}+t_{1}^{2}}{2t_{1}^{2}})^{1/2}~~~~~~~~~~~~~~~|\mu|\leq t_{1},\nonumber~~~~~~~~~~~~~~~(A.11)\label{eq:AEleven}\\
\omega_{-}(q=0)=2t_{1}(1+\frac{g_{\sigma}g_{v}e^{2}t_{1}d}{\kappa v_{F}^{2}}\frac{|\mu|}{t_{1}})^{1/2}~~~~~~~~~~~~~~~|\mu|\geq t_{1},
\nonumber~~~~~~~~~~~~~~~(A.12)\label{eq:ATwelve}
\end{eqnarray}
for the optical plasmon mode in AA-stacked BLG indicating its gaped nature, and $\omega_{-}(q=0)=0$ for the acoustic plasmon mode. One can easily check that Eqs. \ref{eq17}, \ref{eq21} and \ref{eq22} at $q=0$ reduces to the above results, confirming the validity of our results for the low energy dispersion relation of the plasmon modes.

%

%
%
%
\newpage
\begin{figure}
\begin{center}
\includegraphics[width=15cm,angle=0]{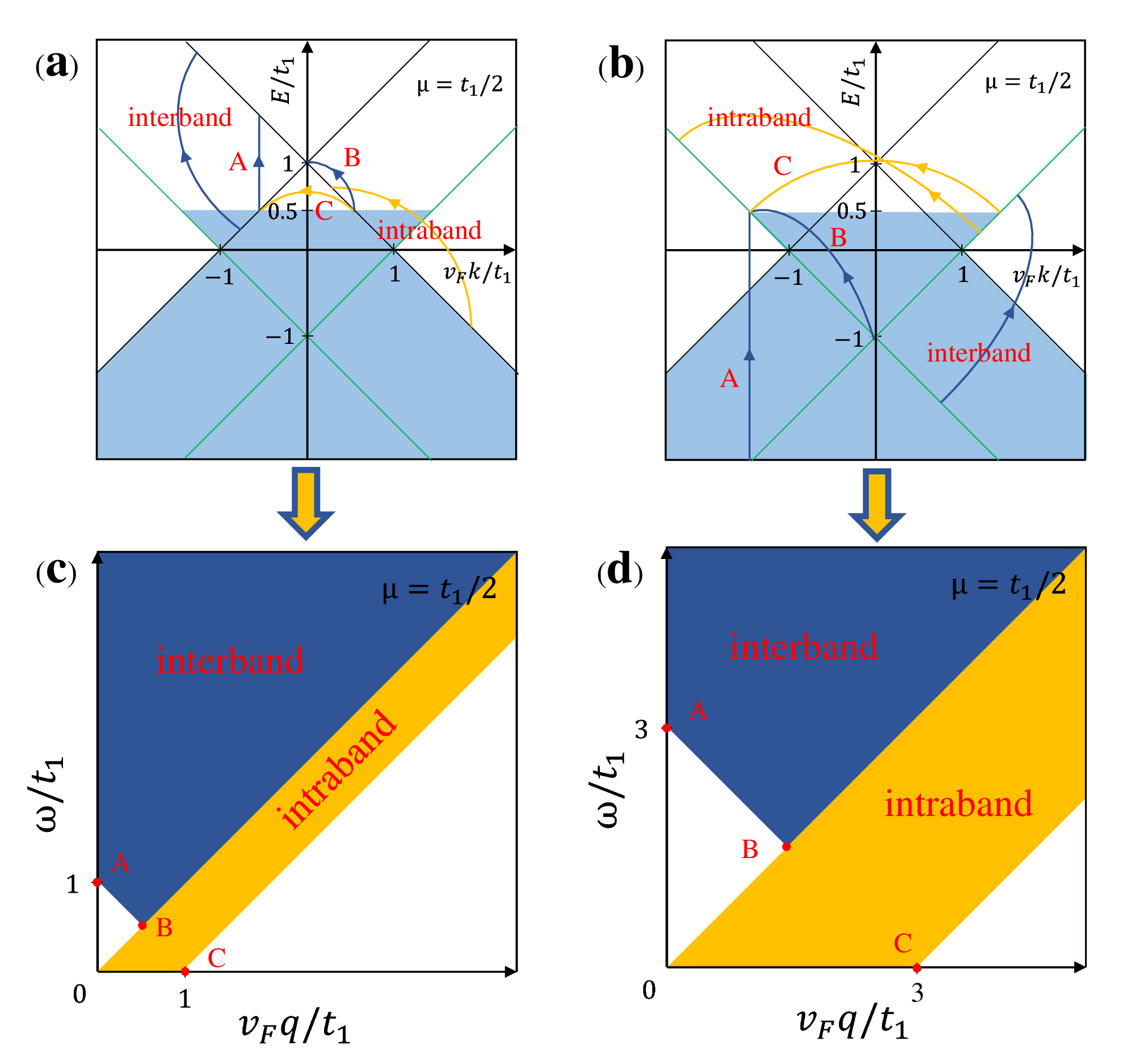}
\caption{The figures in the upper row show all possible intra-pair single-particle excitations for $\mu=t_{1}/2$. Yellow(Blue) lines exhibits the excitations between the band with the same(different) $\lambda$ index, called intra-(inter-)band excitation. The yellow (blue) area of the figures in the lower row show the single-particle continuum corresponding to the intra-pair excitations between the band with the same(different) $\lambda$ index, called intra(inter)band. All borderlines are linear in $v_{F}q$.}
\label{Fig01}
\end{center}
\end{figure}
\begin{figure}
\begin{center}
\includegraphics[width=15cm,angle=0]{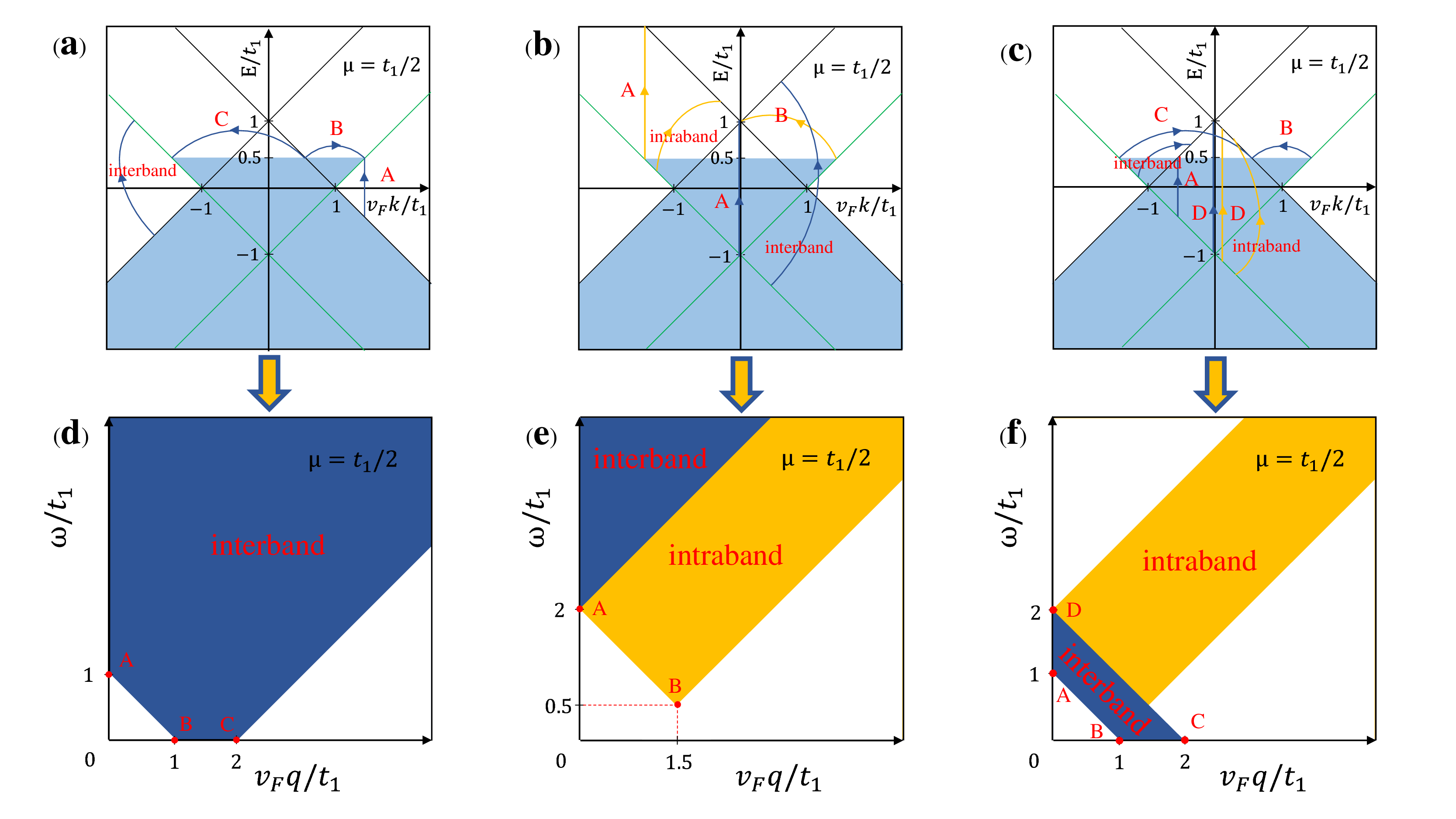}
\caption{The figures in the upper row show all possible inter-pair single particle excitations for $\mu=t_{1}/2$. Yellow(Blue) lines exhibits the intra-pair excitations between the band with the same(different) $\lambda$ index, called intra-(inter-)band excitation. The yellow (blue) area of the figures in the lower row show the single-particle continuum corresponding to the intra-pair excitations between the band with the same(different) $\lambda$ index, called intra(inter)band. All borderlines are linear in $v_{F}q$.}
\label{Fig02}
\end{center}
\end{figure}
\begin{figure}
\begin{center}
\includegraphics[width=15cm,angle=0]{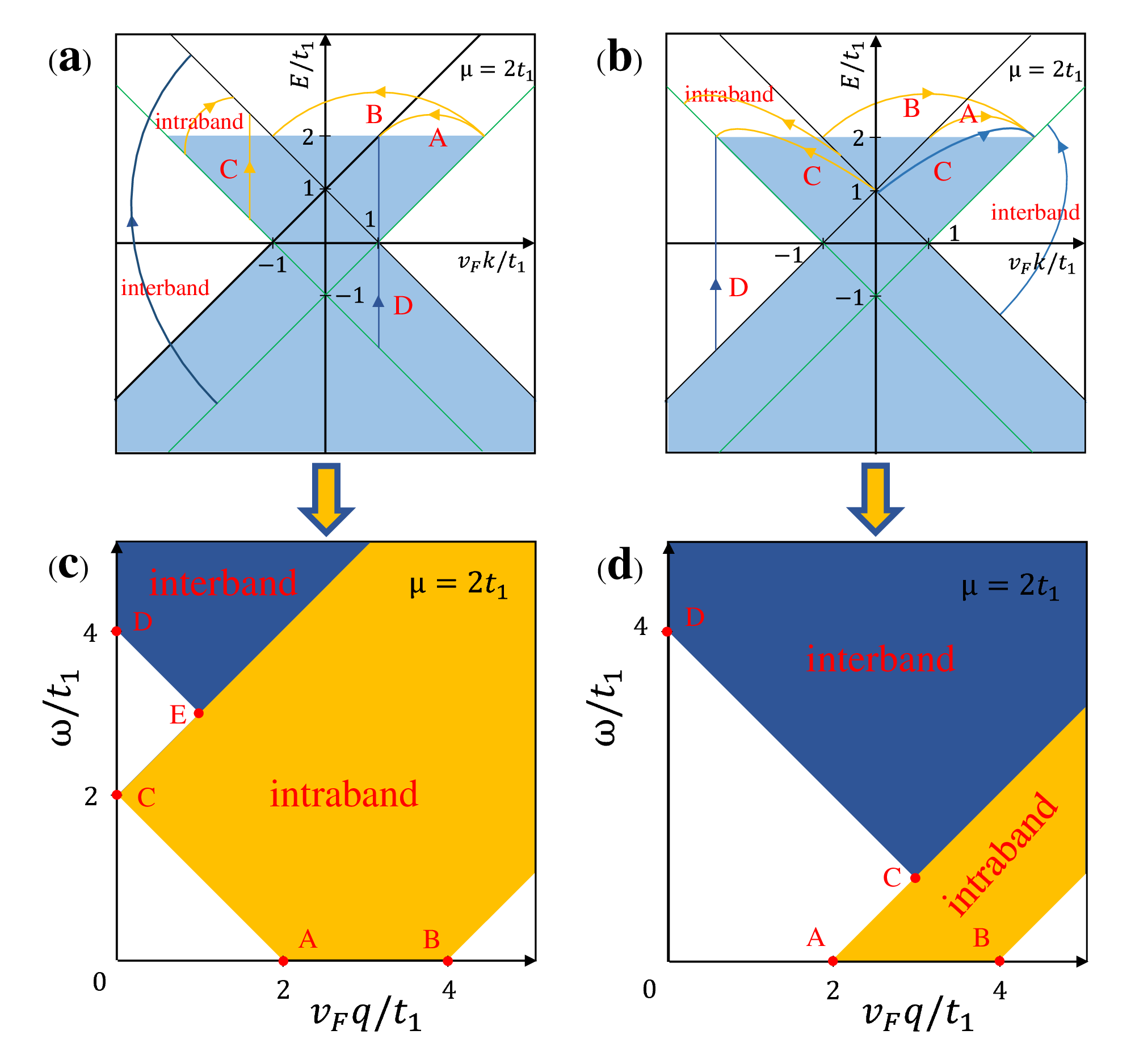}
\caption{Same as Fig. \ref{Fig02} but for $\mu=2t_{1}$}
\label{Fig03}
\end{center}
\end{figure}
\begin{figure}
\begin{center}
\includegraphics[width=15cm,angle=0]{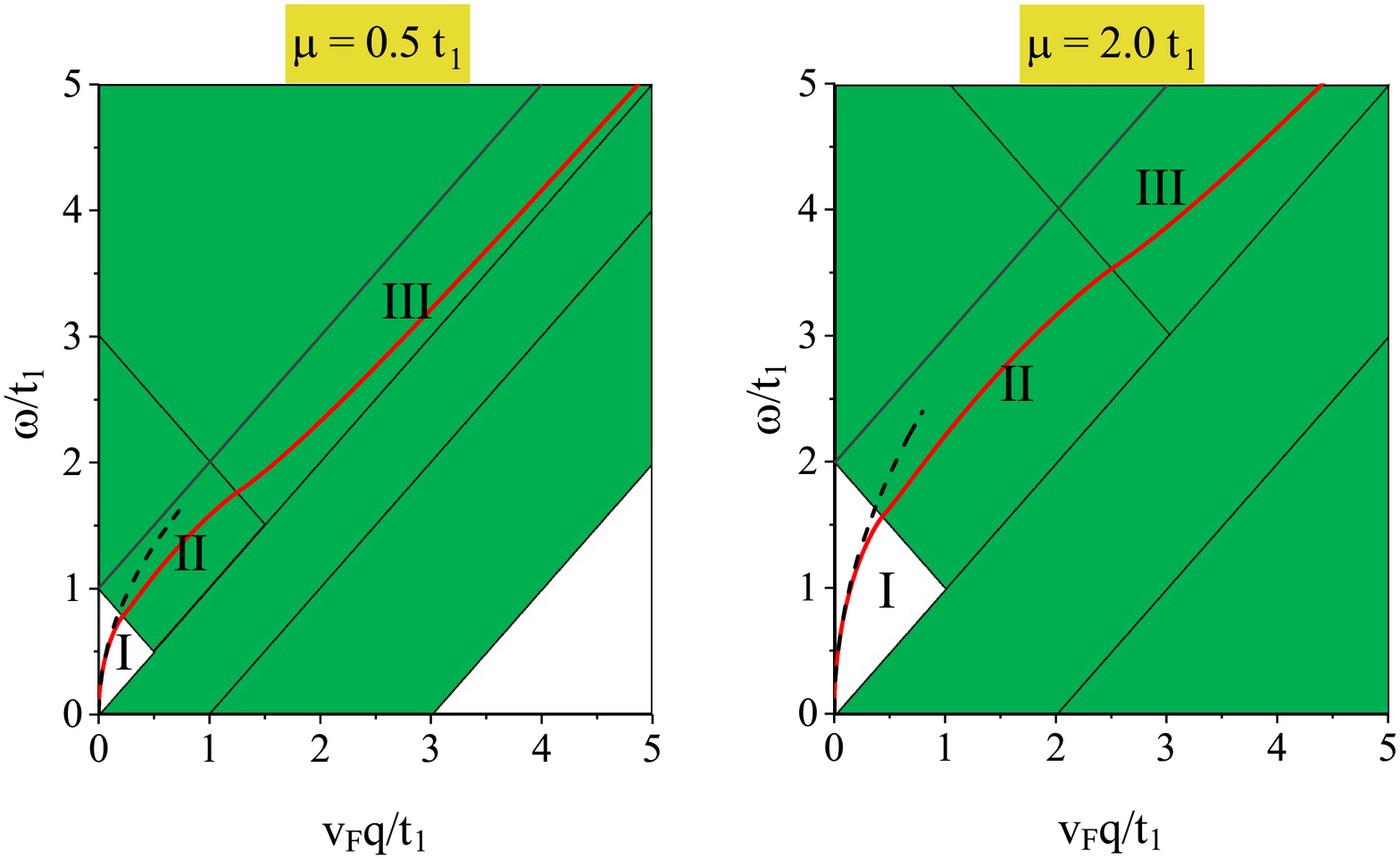}
\caption{Calculated acoustic plasmon dispersion of AA-stacked BLG for different chemical potentials $|\mu|=t_{1}/2$ and $|\mu|=2t_{1}$. The numerical (approximated analytical) dispersion relation of the plasmon  is shown by red solid (black dashed) curve. Green areas show single-particle continuum. All borderlines are linear in $v_{F}q$. The other parameters are $\kappa=2.5$ and $d=3.6$ {\AA}.}\label{Fig04}
\end{center}
\end{figure}
\begin{figure}
\begin{center}
\includegraphics[width=15cm,angle=0]{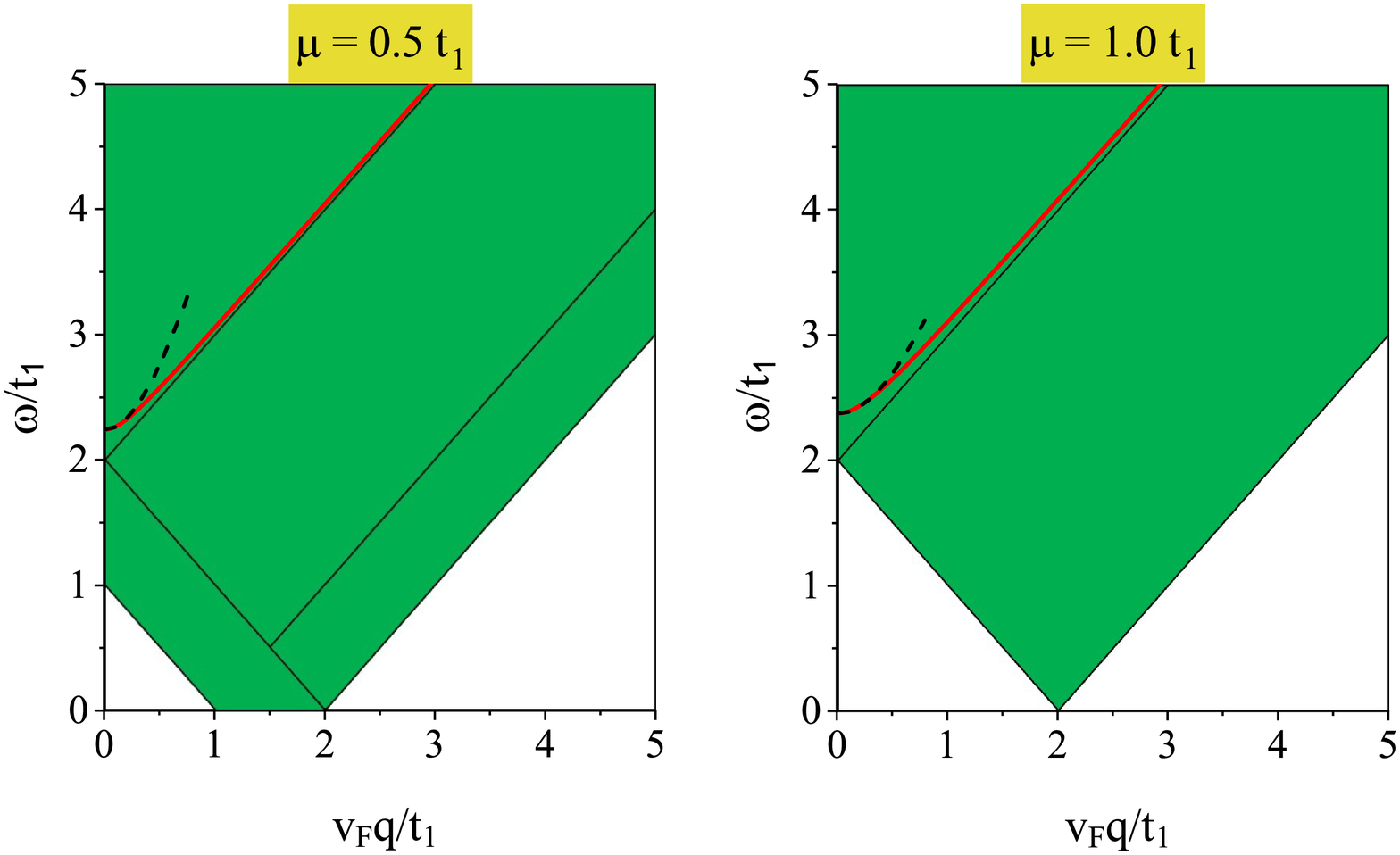}
\caption{Calculated optical plasmon dispersion of AA-stacked BLG for different chemical potentials $|\mu|=t_{1}/2$ and $|\mu|=t_{1}$. The numerical (approximated analytical) dispersion relation of the plasmon  is shown by red solid (black dashed) curve. Green areas show single-particle continuum. All borderlines are linear in $v_{F}q$. The other parameters are $\kappa=2.5$ and $d=3.6$ {\AA}.}\label{Fig05}
\end{center}
\end{figure}
\begin{figure}
\begin{center}
\includegraphics[width=15cm,angle=0]{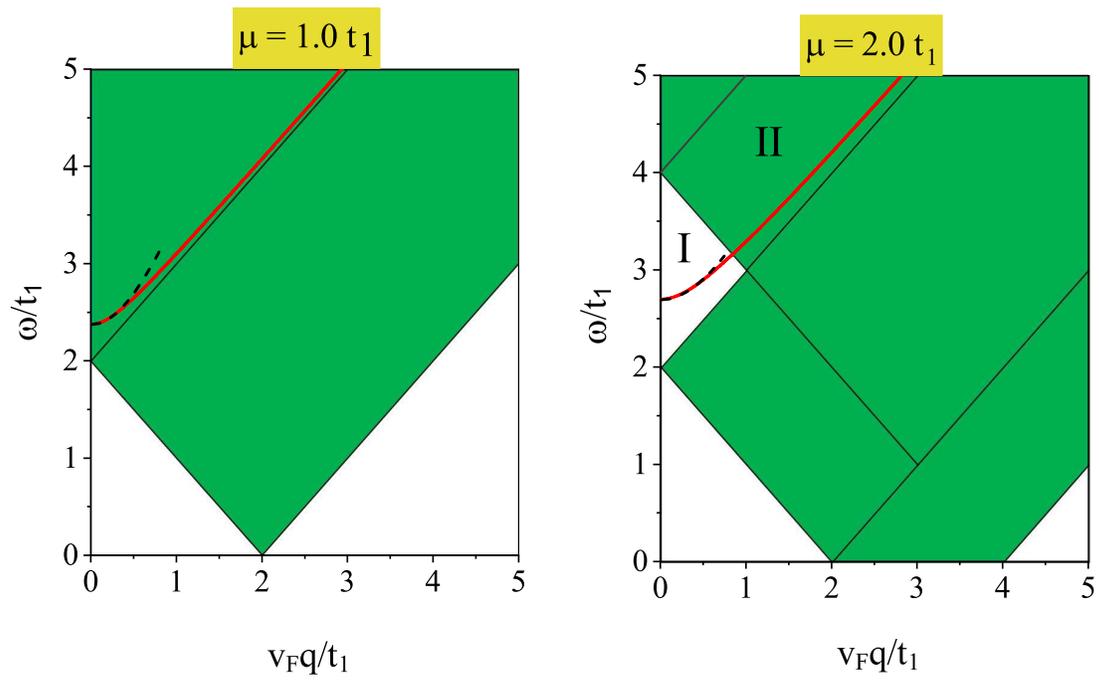}
\caption{Same as Fig. \ref{Fig05} but for $|\mu|=t_{1}$ and $|\mu|=2t_{1}$.}\label{Fig06}
\end{center}
\end{figure}
\end{document}